\documentclass[sigconf]{acmart}
\makeatletter
\renewcommand\@formatdoi[1]{\ignorespaces}
\makeatother

\usepackage{booktabs} 

\setcopyright{none}
\setcopyright{acmcopyright}

\usepackage[inline]{enumitem}
\usepackage{makecell}
\usepackage{fancyhdr}
\usepackage{epsfig}
\usepackage{amsmath}
\usepackage{amsfonts}
\usepackage{color}
\usepackage{algorithmic}
\usepackage{graphicx}
\usepackage{textcomp}
\usepackage{xcolor}
\usepackage{tabularx, ragged2e}

\acmConference[KDD’21 IRS Workshop]{The 27th ACM SIGKDD Conference on
Knowledge Discovery and Data Mining}{August 14-16, 2021}{Singapore}

\copyrightyear{2021}
\acmISBN{}

\begin{document}
\fancyhead{}
\fancyfoot{}

\title{Embedding-based Recommender System for Job to Candidate Matching on Scale}

\author{Jing Zhao, Jingya Wang, Madhav Sigdel, Bopeng Zhang, Phuong Hoang, \newline Mengshu Liu and Mohammed Korayem}
\affiliation{%
  \institution{CareerBuilder}
  \city{Peachtree Corners} 
  \state{Georgia, US} 
  \postcode{30092}
}
\email{{jing.zhao, jingya.wang, madhav.sigdel, bopeng.zhang, phuong.hoang, mengshu.liu, mohammed.korayem}@careerbuilder.com}

\renewcommand{\shortauthors}{Zhao and Wang, et al.}

\begin{abstract}
The online recruitment matching system has been the core technology and service platform in CareerBuilder. One of the major challenges in an online recruitment scenario is to provide good matches between job posts and candidates using a recommender system on the scale. In this paper, we discussed the techniques for applying an embedding-based recommender system for the large scale of job to candidates matching. To learn the comprehensive and effective embedding for job posts and candidates, we have constructed a fused-embedding via different levels of representation learning from raw text, semantic entities and location information. The clusters of fused-embedding of job and candidates are then used to build and train the Faiss index that supports runtime approximate nearest neighbor search for candidate retrieval. After the first stage of candidate retrieval, a second stage reranking model that utilizes other contextual information was used to generate the final matching result. Both offline and online evaluation results indicate a significant improvement of our proposed two-staged embedding-based system in terms of click-through rate (CTR), quality and normalized discounted accumulated gain (nDCG), compared to those obtained from our baseline system. We further described the deployment of the system that supports the million-scale job and candidate matching process at CareerBuilder. The overall improvement of our job to candidate matching system has demonstrated its feasibility and scalability at a major online recruitment site.  
\end{abstract}



\maketitle

\section{Introduction}

\begin{figure}[t]
\centerline{\includegraphics[width=\columnwidth,height=\textheight,keepaspectratio]{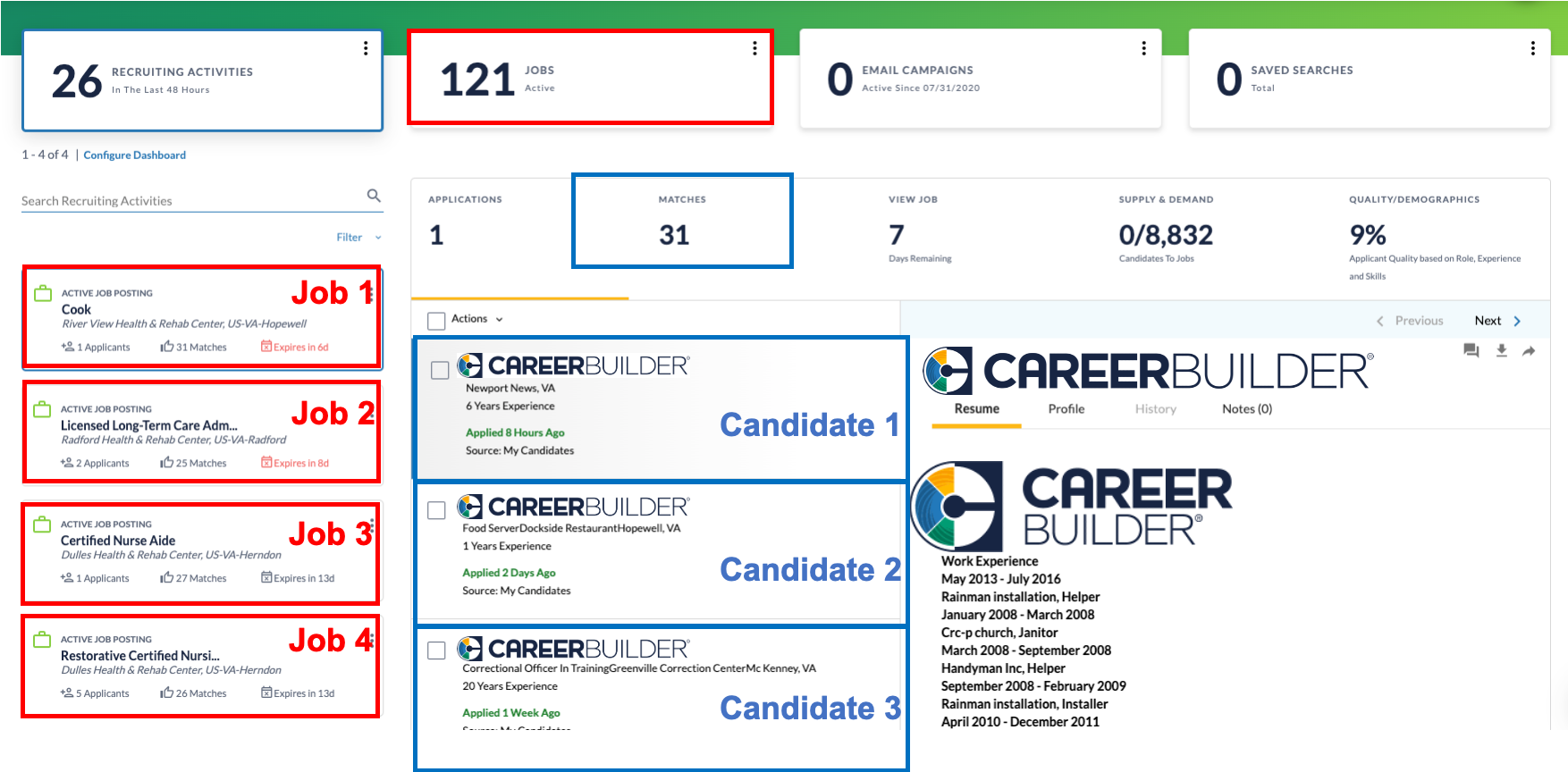}}
\caption{Snapshot of CareerBuilder's online recruitment panel and user interface.}
\label{fig1}
\end{figure}

Efficient and real-time job candidate matching service is not only highly desirable between employers and job seekers, but is also beneficial to the long-term socioeconomic well-being \cite{b1}. The number of both job postings and hiring events through online recruitment platforms has grown rapidly in recent years \cite{b2}. Especially because of the impact of the COVID-19 pandemic, millions of employers and job seekers would prefer to conduct their hiring or job-seeking through the online recruitment platform \cite{b3}. Careerbuilder is the company that possesses the largest online job boards and provides varieties of online recruitment services in the human capital domain. Therefore, the online recruitment matching system has been one of the key services that support CareerBuilder's core business as well as serve millions of customers and users globally. Figure~\ref{fig1} has illustrated a typical job to candidate recommendation scenario that takes place at Careerbuilder every day. The red boxes highlight the posted jobs from the employer and the blue boxes highlight the matched candidates recommended by the algorithm.

With millions of job postings and resumes submitted or updated at CareerBuilder every day, the most critical challenge is to build a recommender system that allows employers to target their fitting candidates and allow the job seeker to find their desired jobs in real-time. To address this challenge, we have proposed a two-staged recommendation system using an embedding-based approach (Figure~\ref{fig3}). A fused embedding strategy that applies deep learning \cite{b4, b5}, representation learning with job-skill information graph \cite{b6} and geolocation calculator \cite{b7} techniques are used for both job and candidate. We have also implemented Faiss index for clustering and compressing the embeddings, which also allows us to conduct the approximate nearest neighbor search for candidate retrieval on runtime \cite{b8, b9}. There are several advantages of using embedding-based recommendation with embeddings. 
\begin{enumerate}
\item Scalability: Easy to scale on the industrial level with embeddings for millions to billions of items with Faiss.
\item Sparsity/Similarity: Content-based embedding provides an alternative way to measure user-item interaction. The pairwise similarity can be easily computed using $l_2$ distance or cosine similarity. 
\item Cold-Start: Mitigate the cold-start issue as this content-based approach does not rely on individual user behavior data.
\end{enumerate}

\begin{figure}[t]
\centerline{\includegraphics[width=0.8\columnwidth,height=\textheight,keepaspectratio]{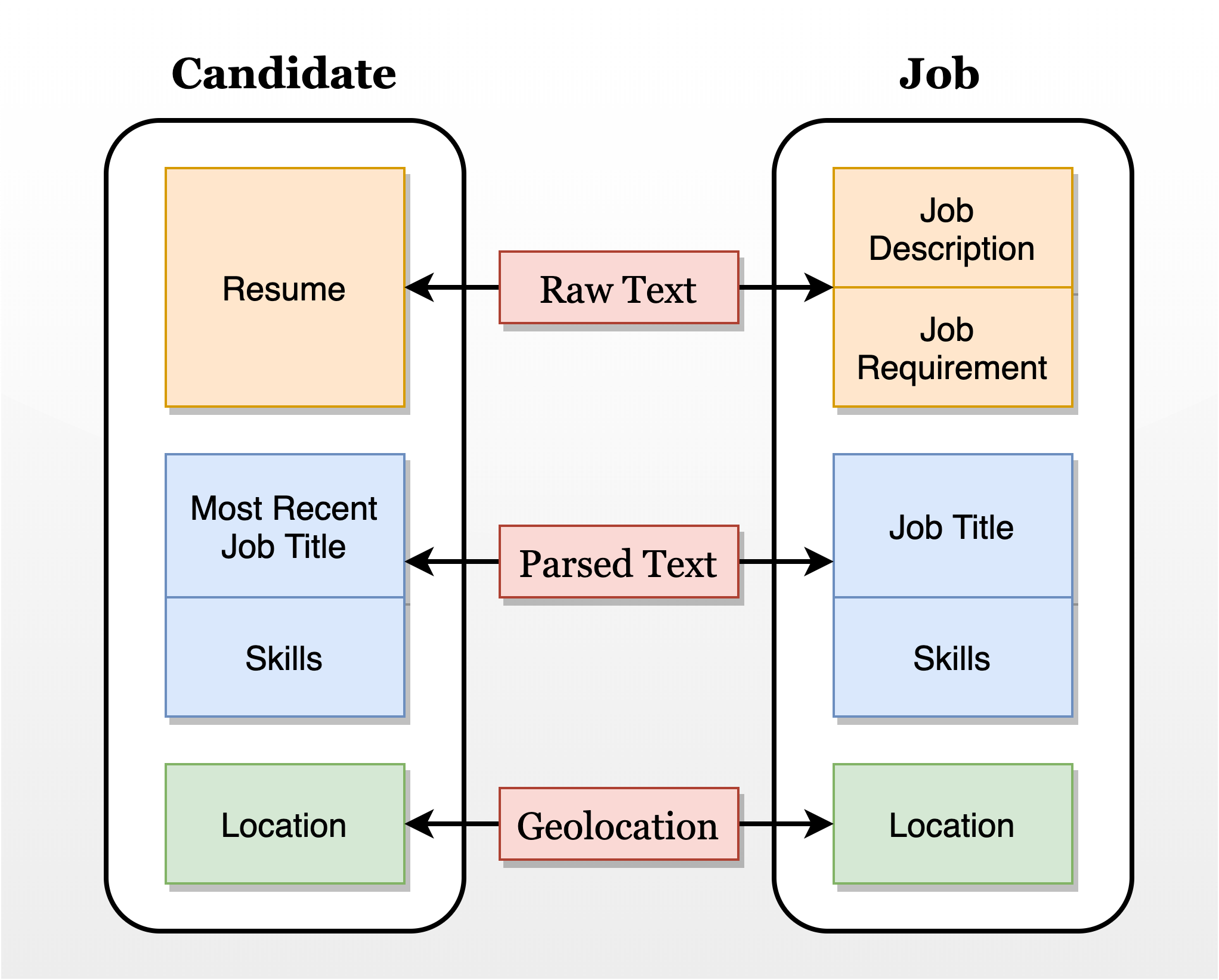}}
\caption{Contextual mapping between job and candidate.}
\label{fig2}
\end{figure}

As for designing the recommender system for online recruitment, a major characteristic that distinguishes it from e-commerce, stream media and social network recommendation scenarios is that the contexts of user and item are likely to be symmetrical. Figure ~\ref{fig2} illustrates such symmetric structure in terms of the context mapping between job and candidate. Active candidates have the motivation to provide full-profile information as it raises their chances to be discovered by the recruiter by search and platform recommendation. At the core of our recommender system, we have taken advantage of such a symmetric structure of contextual mapping to construct a fused embedding using a combination of different strategies. We have applied a convolution neural network (CNN)-based end-to-end approach to learn the effective embedding of the raw text. This deep learning embedding model is equipped with the domain-specific vocabulary to process the text paragraphs from the resume, job description and job requirement. However, deep learning-based models are typically more effective for generalized natural language processing instead of conducting the contextual enrichment for the semantic entity extraction. Therefore, we have also implemented a representation learning model based on the job-skill information graph to parse job title and skill, which includes implicit information of job transition and job-skill co-occurrence that is crucial for the job to candidate matching. Moreover, a geolocation calculator that converts longitude and latitude to three-dimensional Cartesian coordinates is used to construct the location vector. With these three embeddings, we construct a fused embedded representation for both job and candidate by concatenated them together after a weight factor is empirically assigned to each component. 

\begin{figure}[t]
\centerline{\includegraphics[width=\columnwidth,height=\textheight,keepaspectratio]{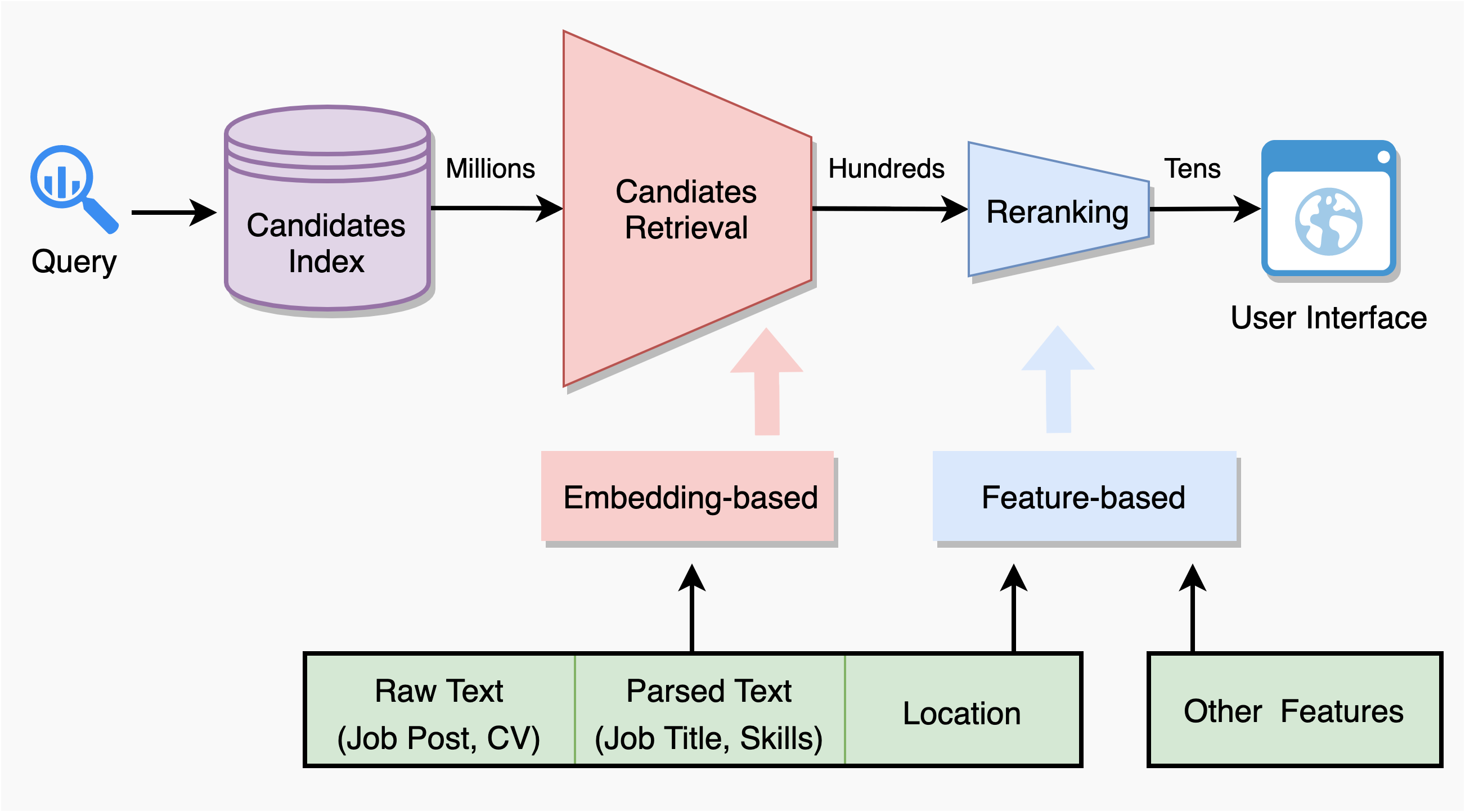}}
\caption{The Architecture of the two-stage online recruitment recommender system.}
\label{fig3}
\end{figure}

\section{Related Works}
\subsection{Recommender System}
Content-based recommender system has the inherent advantages in generalization and mitigating cold-start problems. The content-based embedding strategy allows an easy multi-feature convolution to achieve efficient and reliable item retrieval. Dates back to the classical matrix factorization framework, the content-based features have been incorporated in the recommendation model \cite{b10}. The Factorized Machine can be used as a more generalized model for any content-based feature embeddings \cite{b11}. The rapid development of deep neural networks(DNN) in recent years has opened a new racetrack for developing recommender system. Researchers at YouTube have proposed a recommender system with a Wide \& Deep neural networks architecture \cite{b12}. He \emph{et al} have proposed a neural network based collaborative filtering architecture (NCF) for modeling user-item interactions \cite{b13}. Although Rendle et al. argue that simple dot-product substantially outperforms NCF learned similarities \cite{b14}. The success of recommender system in e-commerce, media and social network has promoted the development of new technologies in this field. For example, knowledge graph has been utilized to build the billion-scale commodity embedding in Alibaba \cite{b15}. Wang \emph{et al}. have also suggested propagating user preferences on the knowledge graph for the recommender system \cite{b16}. As for developing a more dynamical recommender system that also addresses the often delayed logged user feedback, researchers at Google have implemented a policy-gradient-based algorithm that adopted reinforcement learning to build a recommender system \cite{b17}. On the basis of that, a more sophisticated off-policy learning with a two-stage recommendation system is proposed by Ma \emph{et al} \cite{b18}.   

\begin{figure}[t]
\centerline{\includegraphics[width=\columnwidth,height=\textheight,keepaspectratio]{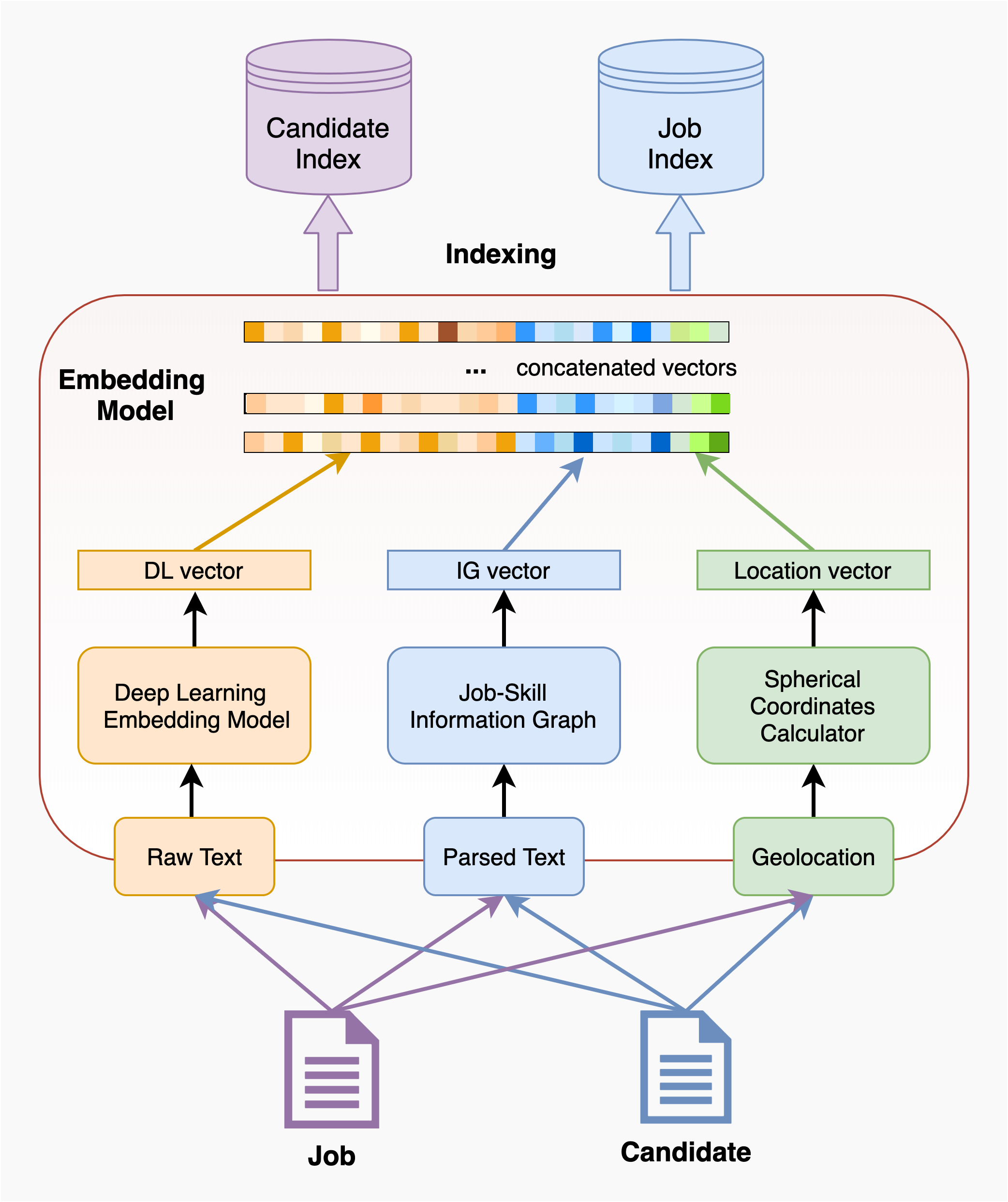}}
\caption[font=bf]{Fused-embedding model with vectors from deep learning embedding model (DLEM), skill-job information graph and geolocation calculator.}
\label{fig4}
\end{figure}

\subsection{Recommender System in the Online Recruitment Domain}
Job and recruitment recommendations in the human capital domain are the particular applications of recommender system that involves text mining, semantic analysis, skill/job title normalization and other NLP techniques. Diaby \emph{et al} proposed a content-based job recommender system along with user's interaction and connection data \cite{b19}. Rafter \emph{et al} proposed a user-based collaborative filtering (CF) system that utilizes the overlapping of interacted jobs as the similarity measure between two users. They have also applied a nearest neighbor search approach to generate recommendations \cite{b20}. To overcome the sparsity and cold-start problems of the classical CF method, Shalaby \emph{et al} has a scalable item-based recommendation system by leveraging a directed graph of job connection to represent the user behavior and contextual similarity \cite{b21}. Bian \emph{et al} has proposed a deep global match network for capturing the global semantic interactions between job posting and candidate resume at both sentence and global levels \cite{b22}. Jiang \emph{et al} proposed using deep learning and LSTM to learn the explicit and implicit interaction between job and candidate to get a more comprehensive and effective representation for the matching \cite{b23}.

\section{Candidate Matching System and Architecture}
The proposed architecture of the two-staged recommendation system consists of two major components (Figure~\ref{fig4}): 
\begin{enumerate}
\item First stage retrieval component that utilizes two-tower embedding structure to find hundreds of potential candidates from the pool of millions.
\item Second stage rerank component that takes advantage of various contextual features allows the narrow down to a few dozen of candidates after the fine-tune scoring.
\end{enumerate}
At the core of the first component, we have proposed a fused embedding strategy to learn the representations from raw text, parsed text and geolocation for both candidate and job. 

\begin{figure*}[t]
\centerline{\includegraphics[width=0.9\textwidth,height=0.9\textheight,keepaspectratio]{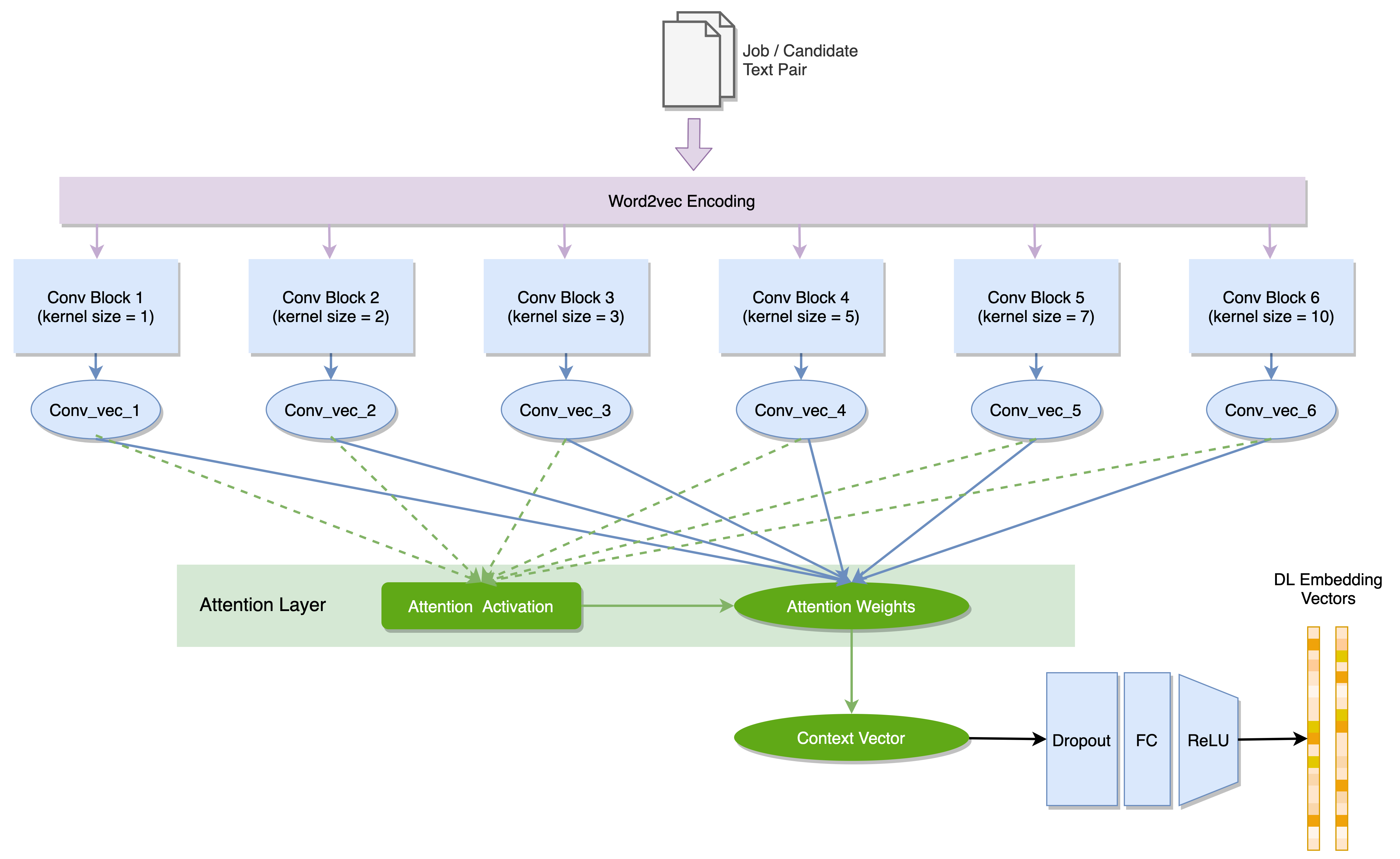}}
\caption{Architecture of the deep learning embedding model (DLEM).}
\label{fig5}
\end{figure*}

\subsection{Deep Learning Embedding Model}\label{AA}
We have trained an end-to-end Deep Learning Embedding Model (DLEM) on a supervised learning task that utilizes our job application data. This allows the DLEM model not only learns the context embedding from an NLP perspective but also being able to capture the job application behavior from the users. The DLEM consists of an input layer, a convolutional neural network (CNN) layer and an attention layer as illustrated in Figure~\ref{fig5}. At the data generation stage, a pair of job and candidate's raw text documents (e.g. job posts and resume) are generated for the input layer. The positive pairs are particularly selected from our job application logs in which the candidate is paired with the job that he/she applied for. The negative pair is generated using random samples but the results are filtered with additional rules to remove the false negative signals. For example, the job and candidate pair that belong to the same SOC domain are removed from the negative samples. The pairwise raw text inputs are then encoded using word2vec using a domain-specific vocabulary with a focus on the human resource and job domain. With this domain-specific encoding of the word index, we are able to construct a more space-efficient index-based representation. The input text encoding is then sent to the convolutional layer, which consists of six stacked blocks with different kernel sizes, ranged from 1 to 10. Each stacked block contains three consecutive convolutional blocks, in which a pipeline of 1D-convolution, batch normalization and max-pooling is considered as a unit processing. The stacked blocks with different kernel sizes are aimed to construct the distributed representations of the sentence instead of just the lexical features. An attention layer is built from the outputs from the stacked blocks and their saliency, inspired by the recent progress of the Transformer architecture \cite{b24}. The output context vector of the attention layer is then sent to the fully-connected layers (FC layers) with RELU activation. FC layers also determine the desired output dimension of the embedding vector based on the need. As for training the DLEM, we have chosen a relevance-based binary cross-entropy as the loss function. 

\(C\) and \(J\) represents the sets of candidates and jobs. The application mapping $A$ is defined as \(A: C \rightarrow 2^{J}\), and relevancy mapping $R$ is defined as \(R: C \times J \rightarrow \{0, 1\}\)
\begin{align*}
    \text{such that} \quad
    R(c, j) &=
    \begin{cases}
    1,& \text{if} ~ c ~ \text{is relevant to} ~ j\\
    0,              & \text{otherwise}
    \end{cases}.
\end{align*}
\(g(x)\) is defined as the embedding from DLEM model and \([g(c);g(j)]\) represents the concatenation of candidate embedding vector \(g(c)\) and job embedding vector \(g(j)\).

\begin{equation*}
\begin{aligned}
& Loss(C, J) = \\
& \\
&\sum_{c\in C}\sum_{j\in A(c)} - R(c, j)\log(p) + (1 - R(c, j))\log(1-p)  \\
& \\
& \textrm{where} \;\;\; p = \sigma\left(w^T[g(c);g(j)] + b\right) \\
\end{aligned}
\end{equation*}

\begin{figure*}[t]
\centerline{\includegraphics[width=\textwidth,height=\textheight,keepaspectratio]{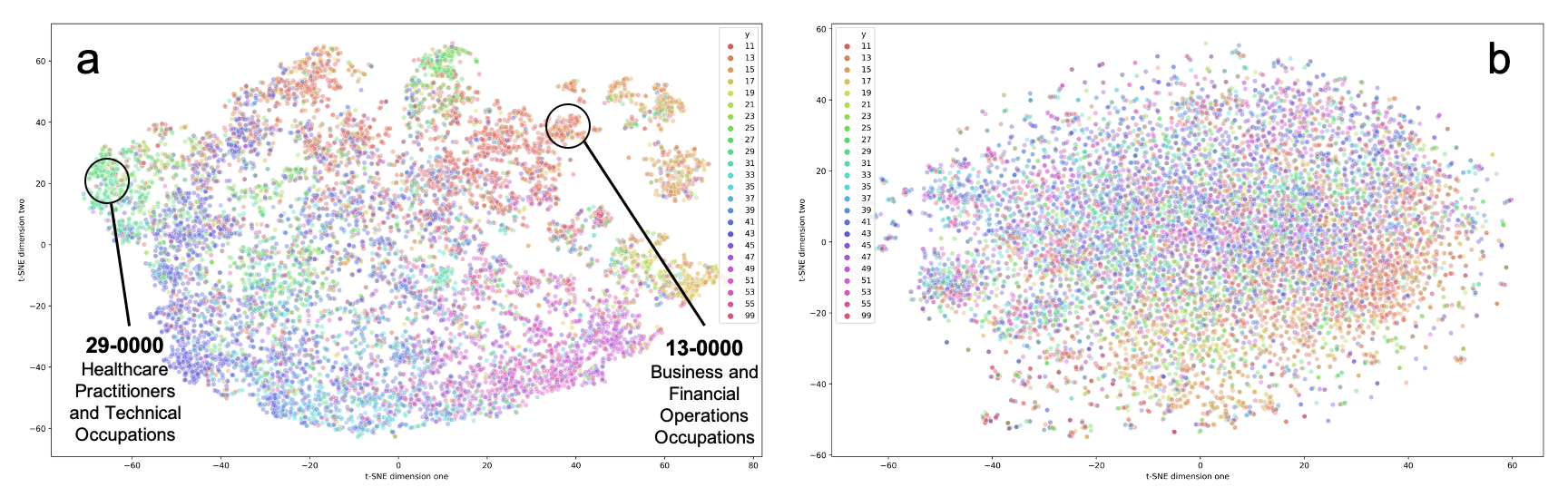}}
\caption{t-SNE plot of job embedding vectors from (a) DLEM and (b) distilBERT that are color labeled with 24 SOC categories.}
\label{fig6}
\end{figure*}

Figure~\ref{fig6} illustrates the t-distributed stochastic neighbor embedding (t-SNE) plots of 10,000 sample jobs' embeddings obtained from (a) DLEM and (b) distilBERT pre-trained model \cite{b22}. Each job is also color labeled with 23 major job categories and one unknown category based on the Standard Occupational Classification System (SOC). The t-SNE plot shows that our DLEM is very effective in job classification as the job cohorts with different colors are clearly clustered in different regions. For example, job category 29-0000, Healthcare Practitioners and Technical Occupations and 13-0000 Business and Financial Operation Occupations have their distinguished clustering circled on the plot. For comparison, we cannot observe a structural clustering of the embeddings obtained from the pre-trained distilBERT model. This might due to the lack of a specific domain dictionary and labeled training data for the distilBERT model.  

\subsection{Representation Learning with Job-Skill Information Graph}
Job title and skill are considered the most important semantic entities as they are (semi-)structured fields and contain enriched information in the job-related documents. Traditionally, semantic matching using job title and skill entities has been the focuses for job classification and job recommendation tasks. Herein, we have taken advantage of a representation learning model that utilizes the information graph from job transition network, job-skill network and skill co-occurrence network \cite{b6}. The model used both Bayesian personalized ranking and margin-based loss functions to learn the vector representation for the semantic entities and allow us to encode the local neighborhood structures captured by the information graphs. The following three objective functions are used to learn the representation for \(W\) and \(W^\prime\), which correspond to the representation of job title and skill, respectively.  

\begin{equation*}
\begin{aligned}
O^{jj} &= \min_{W} - \sum_{(x, y, z)\in \mathcal{D}^{jj}} \ln\sigma\big(\langle \mathbf{w_x}\cdot\mathbf{w_y}\rangle - \langle\mathbf{w_x}\cdot\mathbf{w_z}\rangle\big) \\
O^{ss} &= \min_{W^\prime} - \sum_{(x, y, z)\in \mathcal{D}^{ss}} \ln\sigma\big(\langle \mathbf{w_x^\prime}\cdot\mathbf{w_y^\prime}\rangle - \langle\mathbf{w_x^\prime}\cdot\mathbf{w_z^\prime}\rangle\big) \\
O^{js} &= \min_{W,W^\prime} - \sum_{(x^j, y^s, z^s)\in \mathcal{D}^{js}} \ln\sigma\big(\langle \mathbf{w_{x^j}}\cdot\mathbf{w_{y^s}^\prime}\rangle - \langle\mathbf{w_{x^j}}\cdot\mathbf{w_{z^s}^\prime}\rangle\big)
\end{aligned}
\end{equation*}

Where \(\mathcal{D}^{jj}\) represents the transition relationship of job triplets \((x, y, z)\),  \(\mathcal{D}^{ss}\) represents the co-occurrence of skill triplets and \(\mathcal{D}^{js}\) represents the relationship between (job, skill, skill) triplets. \(\langle\mathbf{w_x}\cdot\mathbf{w_y}\rangle\) is the dot product of two embeddings, which is then used as the input for the sigmoid function \(\sigma(x)\) to calculate the probability. 

\begin{equation*}
O\left(\mathbf{W}, \mathbf{W^\prime}\right) = \min_{W,W^\prime} O^{jj} + O^{js} + O^{ss} + \lambda\left(\|\mathbf{W}\|_F^2 + \|\mathbf{W^\prime}\|_F^2 \right)
\end{equation*}
To unify these three types of networks between job and skill, the joint objective function with \(l_2\) normalization is applied to avoid the over-fitting of representation \(W\) and \(W^\prime\). 

\subsection{Geolocation Calculator}
In the geolocation part, we have calculated the spherical coordinates representation of latitude \(\theta\) and longitude \(\phi\) to the Cartesian coordinates \([x, y, z]\) using the following equations:
\begin{equation*}
\begin{aligned}
& x = \cos(\theta)\cos(\phi) \\
& y = \cos(\theta)\sin(\phi) \\
& z = \sin(\theta) 
\end{aligned}
\end{equation*}
The Cartesian coordinates location vector \(c = [x, y, z]\) has a straightforward advantage for conducting dot product operations between two vectors. The larger dot product \(\langle \mathbf{c_1}\cdot \mathbf{c_2} \big \rangle\) between the location vectors, the shorter distance between these two locations. This relationship is revealed by the following equation:

\begin{equation*}
 d = 2r\cdot\arcsin\left(\frac{\sqrt{r^2 - 2\big \langle \mathbf{c_1}\cdot \mathbf{c_2} \big \rangle}}{2}\right) 
\end{equation*}

in which \(r\) is the radius of the earth and \(d\) is the great-circle distance between two locations on the earth. Therefore, it has the same property as the content-based embeddings when compares to the pair-wise similarity using dot product operation. So the Cartesian location vector is incorporated in the fused embeddings as well.

\subsection{Approximate Nearest Neighbor Search}
The embeddings from DLEM \(\mathbf{v_{dlem}}\), job-skill information graph \(\mathbf{v_{ig}}\) and geolocation calculator \(\mathbf{v_{geo}}\) are concatenated together with a set of empirically assigned weights for with each component. The concatenated embedding \(\mathbf{v_{fused}}\) are defined as:
\begin{equation*}
\left[\mathbf{w_{1}\cdot v_{dlem}^\intercal};  \mathbf{w_{2}\cdot v_{ig}^\intercal};  \mathbf{w_{3}\cdot v_{geo}^\intercal}\right] \rightarrow \mathbf{v_{fused}^\intercal}
\end{equation*}

After constructing the fused embedding vectors, we employed the Faiss index to store all of our item embeddings for search and retrieval. This brings several advantages:
\begin{enumerate}
\item Faiss index requires less space for storage due to product quantization of the embedding vectors \cite{b23}, which is essential for both our offline spark pipeline and online services that possess tight memory restriction.
\item It is easy to be integrated into the system for item retrieval. The inverted file index (IVF) allows a runtime approximate nearest neighbor search from millions or even billions of items.
\item We can easily evaluate the similarity score between job and retrieved candidates using the inner product or $l_2$ metric from the index. 
\end{enumerate}
There are several factors we have considered during the customization of the Faiss index. 1. We have chosen IVF algorithms and carefully tune the number of coarse clusters during the coarse quantization, which typically works through the K-means clustering; 2. As for the fine-grained quantization, we have applied OPQ to transform data prior to the product quantization, which is recommended by Huang \emph{et al} \cite{b9}; 3. We have also tuned the \emph{nprobe} parameter that decides how many coarse clusters will be scanned during the query, which may affect the retrieval's performance and recall. Overall, the architecture of both job and candidate index resembles the two-tower model, which has demonstrated its effectiveness in text-based information retrieval in large-scale recommender system \cite{b24, b25}.   

\subsection{Reranking with Contextual Features}
After the first stage candidate retrieval, the final ranking score for each candidate is calculated by a weighted linear equation that aggregates the scores we obtained from the first-stage relevancy score as well scores from contextual features of job and candidates. These context-based scores include skill matching, location restriction, year of experience and education level. The weights representing the importance of each score and are tuned empirically. The final ranking score is then used for reranking to generate the second-stage recommendation result. The fine-tuning in the reranking stage also allows us to implement some specialization for a certain type of job. Since the pandemic, there is a significant increase amount of Work From Home (WFH) or remote jobs that appeared in the job posts \cite{b29}. This type of job typically has very little or no location restriction, which is distinguished from a lot of front-line occupations. To reflect such distinction in our recommendation result, we can adjust the location weight during the reranking, which resulted in a more suited and robust candidate recommendation overall.

\begin{table*}[htp]
\caption{Case Study of Job to Candidate Matching Scenarios}
\begin{center}
\setlength{\extrarowheight}{1pt}
\begin{tabularx}{0.9\textwidth} { 
  | >{\centering\arraybackslash}c 
  | >{\centering\arraybackslash}c 
  | >{\RaggedRight\arraybackslash}X 
  | >{\centering\arraybackslash}c 
  | >{\RaggedRight\arraybackslash}X | }
 \hline
 \textbf{Case} & \multicolumn{2}{|c|}{\textbf{Job}} & \multicolumn{2}{|c|}{\textbf{Candidate}} \\ [2pt]
 \hline
  & Title & Database Developer & Previous Title & Sr. Database Developer \\ \cline{2-5}
 1 & Requirement & BS in Computer Science, 5+ years of experience working with Microsoft SQL Server, database code development,  data modeling with ER/Studio or ERWin,  data warehouse design,  SSIS & Skills & Web development and database architecture, Microsoft SQL Servers, Java, C\#, Visual Basic, Microsoft Access  \\[24pt] \cline{2-5}
  & Description &  Design and development of database objects, populate and maintain the data in the data warehouse,  creation of ETL programs in a Microsoft SQL Server environment.  & Work Experience & \begin{itemize}
  \item Sr. Database Software Developer (13 - 20)
  \item  Sr. Database Developer (09 - 13)
  \item Sr. Web Program Developer (02 - 09)
\end{itemize} \\[24pt] \cline{2-5}
  & Location & San Diego, CA & Location & Beaumont, CA  \\
 \hline
  & Title & Licensed Practical Nurse (LPN) & Previous Title & Licensed Practice Nurse \\ \cline{2-5}
  2 & Requirement & Current LPN license in good standing, CPR certification, Minimum 1 year clinical experience. & Skills & LPN, CPR BLS certified, nursing care practice, physical examination, IV drug therapy management, EMR systems \\[16pt] \cline{2-5} 
  & Description & Client assessment, administration of prescribed medication, treatment and therapy, clinical works, supply management, emergency Management. & Work Experience & \begin{itemize}
  \item Licensed Practical Nurse Endoscopy (19 - 20)
  \item LPN Charge Nurse (16 - 19)
  \item State-tested Nurse Assistant (09 - 15)
\end{itemize}\\[20pt] \cline{2-5} 
   & Location & Carnegie, PA & Location & Bulter, PA  \\
 \hline
  & Title & Regional Sales Representative & Previous Title & Regional Sales Representative \\ \cline{2-5}
  3 & Requirement & Require 5+ years of outside sales experience, ability to travel up to 50\% of the time, interpersonal communication skills, experience with CRM platforms, proficient in Microsoft Office & Skills & Client relationship management, communication and negotiation, proficient in salesforce and other CRM platforms, analytical skills, Microsoft Office \\[24pt] \cline{2-5} 
  & Description & Generate, develop, and maintain a robust pipeline of qualified opportunities, actively conduct cold and warm calling to prospective leads, manage sales process.  & Work Experience & \begin{itemize}
\item Regional Sales Representative (19-20)
\item  Business Development Specialist   (18 - 19)
  \item Account Manager II (14-18)
\end{itemize}\\[18pt] \cline{2-5} 
 & Location & North America & Location & Bedford, TX  \\
\hline
\end{tabularx}
\label{tab1}
\end{center}
\end{table*}

\subsection{System Implementation}
The job and candidate data are stored in our in-house Hadoop clusters which allows distributed processing using Spark. The deep learning model is served in the spark jobs to create document embeddings. The fused-embeddings are then used to train the Faiss index with coarse quantization and product quantization (PQ). The published inverted file (IVF) Faiss index is then served for the candidate retrieval in the batch offline mode. All the spark jobs are scheduled by the Oozie coordinator that runs periodically. At the end of the workflow, the generated recommendation results are delivered to the production database.

\section{Experiment And Result}
The test and evaluation of our job to candidate matching system has taken advantage of a rich corpus of job and candidate data at CareerBuiler.com. CareerBuilder operates the largest job posting board in the U.S. and has quickly expanded its global presence in recent years. On the daily routine, millions of job postings and more than 60 million actively searchable resume needs to be processed for the online recruitment service. In this section, we described the details of the case study, testing and evaluation of our system. 

\subsection{Case Study of Matching Scenarios}
The two-stage job-to-candidate matching system has achieved impressive matching quality which is showcased in the table. Table~\ref{tab1} has presented 3 cases with jobs and their top candidates. Each job has its job title, job requirement, job description and location information. The corresponding information from the candidate, such as most recent title, skills, work experience and location are provided as well. As for case 1, the database developer job, the top candidate has shown matching for all four aspects. As for case 2, the licensed practical nurse (LPN) job. We noticed that top candidates meet the requirement for LPN license and other required certificates. Case 3 regional sales representatives job does not have a specific location but north American region, therefore a broader spectrum of candidates can be selected as long as it meets the location requirement. This case also applies to work from home job scenario in which the candidate's working location is not restricted.  Overall, our job-to-candidate matching system has provided satisfied matching results from title, description, requirement and location perspectives, which indicates the success of our two-staged model and fused-embedding strategy.

\subsection{Offline Evaluation}
The DLEM cutoff parameter \(\gamma\), fused embedding weight parameters \(w_{1}\), \(w_{2}\), \(w_{3}\) and score aggregation parameters during the heuristic re-ranking were all tuned empirically through multiples rounds of test and evaluation. QA team and professional recruiters at CareerBuilder also participated in the qualitative evaluations for several rounds. They are asked to validate the list of recommended candidates from both jobs in specific domains and randomly sampled jobs. They give the qualitative score and leave comments for each job-candidate pair. The feedback has been used as the empirical signal for us to better tweak the parameters and search for the optimal parameter combinations for our system. After the fine-tuning of the parameters, we compared the quality score and nDCG between our baseline model and our two-staged matching model. For background information, our baseline model is a solr-powered recommendation engine that utilizes hierarchical classification and a content-based approach to retrieve relevant candidate profiles. As for the offline evaluation, 150 jobs that spanning over multiple job categories with ~3k matching candidates are manually examined. The overall quality score of recommendation has improved $\sim$19\%, the nDCG has improved $\sim$18\% (Figure ~\ref{fig7}). 

\subsection{Online Evaluation}
\begin{figure}[t]
\centerline{\includegraphics[width=1.0\columnwidth,height=\textheight,keepaspectratio]{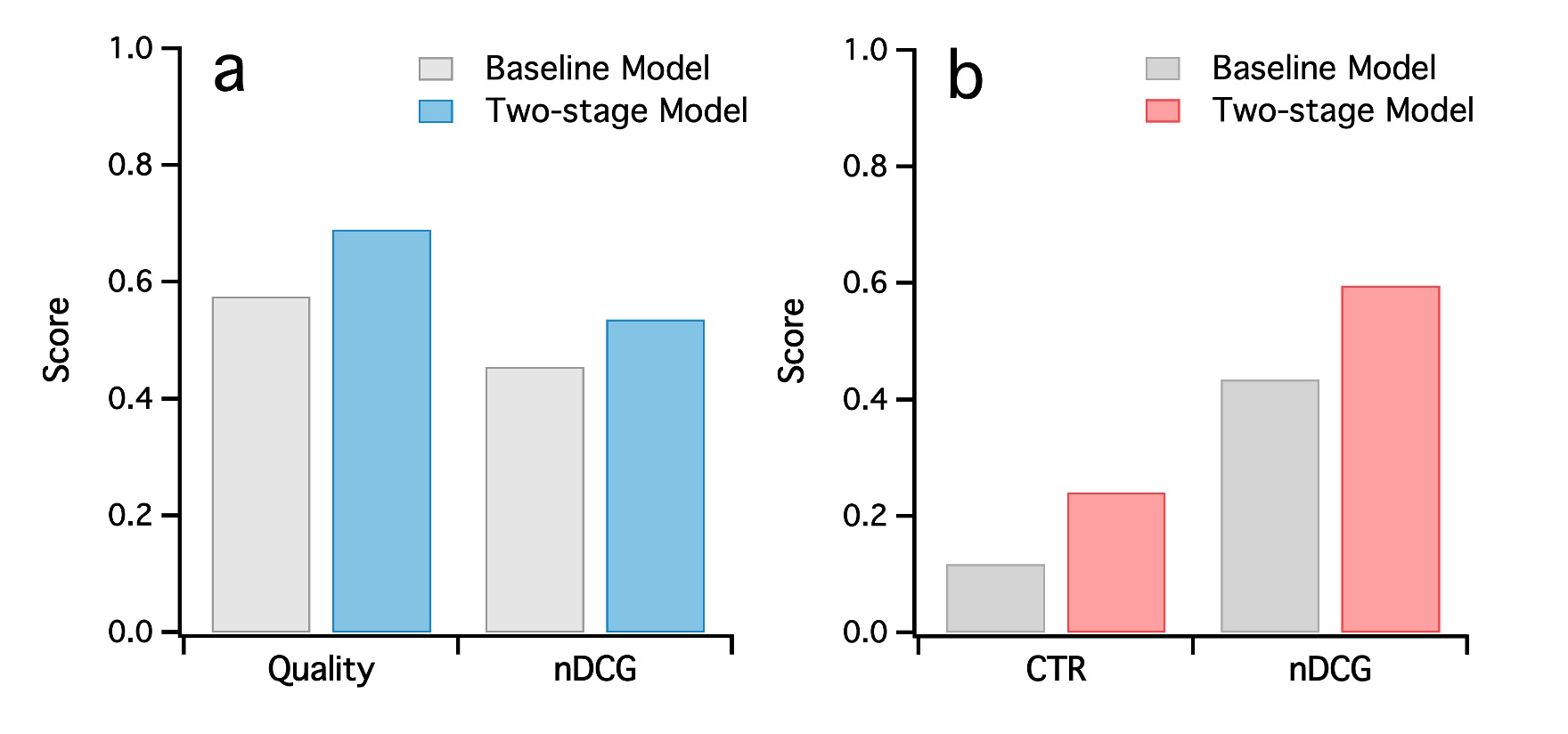}}
\caption{Offline and online evaluation results between baseline model and two-stage matching system.}
\label{fig7}
\end{figure}

As for the online evaluation, we have compared the traffics over 4 months between the baseline model and two-stage matching system. Over ~120k user's impression and click events have been used to calculate the nDCG and click through rate (CTR) for comparison. The CTR and nDCG have both shown significant improvement over three months period of time. The CTR has increased $\sim$104\%, and nDCG has increased $\sim$37\%. These results have also been summarized in Figure ~\ref{fig7}. In summary, both offline and online evaluation results suggest that our two-stage matching system has significantly improved the matching quality, resulted in higher traffic and CTR from our users. 

\section{Conclusion}
The online recommender system has gained considerable attention in both academia and industry in recent years as quickly evolved technology plays a key role in bringing an enormous amount of commercial and social values. The online recruitment service at CareerBuilder has also taken advantage of such progress to serve millions of job applicants and employers. To bring the full potential of the recommender system for online recruitment, we have proposed a two-stage embedding-based recommender system for job to candidates matching. The architecture of this system consists of a two-stage recommendation procedure, a fused-embedding component for candidate retrieval and a fine-tuning reranking module. The successful deployment of embedding-based job to candidate matching system in production creates the avenue to optimize the system end to end through the users' feedback. We also introduced valuable experience in architecture design, serving algorithms parameter tuning and later-stage optimization. Overall, our two-stage job to candidate matching system has shown a significant improvement over the baseline model by measures of CTR and nDCG in real world production environment, which provides an excellent example for deploying an embedding-based recommender system for applications of job to candidate matching on the scale. 

\section*{Acknowledgments}
The authors would like to pay special tribute to Bopeng, who has sadly passed away during the drafting of this paper. We would also like to dedicate this paper to Bopeng to recognize his crucial contribution and achievement during his days at CareerBuilder.

\end{document}